\documentstyle[12pt,oldlfont]{article}

\newlength{\aivwidth}   
\newlength{\tmpwidth}   

\newlength{\zeichen}
\newlength{\strich}
\begin{document}
\begin{titlepage}
\title{The Equivalence Theorem\\for the Heavy-Higgs Standard Model\\
and the Gauged Nonlinear $\sigma$-Model}
\author{Carsten Grosse-Knetter\thanks{E-Mail:
knetter@physw.uni-bielefeld.de}\\[5mm]Universit\"at Bielefeld\\
Fakult\"at f\"ur Physik\\Postfach 10 01 31\\33501 Bielefeld\\Germany}
\date{BI-TP 94/25\\hep-ph/9405412\\May 1994}
\maketitle
\thispagestyle{empty}
\begin{abstract}
The equivalence theorem states that the leading part of the amplitude
for a process with external longitudinally polarized vector bosons
is given by the amplitude in which the longitudinal vector bosons are
replaced by the corresponding pseudo-Goldstone bosons. The validity
of this theorem within the standard model with a heavy Higgs boson
and within
the gauged nonlinear $\sigma$-model (in which the Higgs boson
is absent) is shown. Furthermore it is examined to what
extent also internal lines other than scalar lines can be neglected.
A simple power-counting
method is developed which determines the leading diagrams
for a given process at an arbitrary loop order. This method is also
applied to effective Lagrangians with additional nonstandard
interaction terms of higher dimension (chiral Lagragians).
\end{abstract}
\end{titlepage}

\section{Introduction}
\typeout{Section 1}
The equivalence theorem (ET) \cite{coleti,lequth,chga,gokone,ve}
simplifies calculations of $S$-matrix elements
for scattering processes with external longitudinally polarized
massive vector bosons at high energies ($E\gg M_W$).
This theorem states that the leading part of
the amplitude for such a process is equal to the amplitude
in which
the longitudinally polarized vector bosons are replaced by the
corresponding unphysical Goldstone bosons (calculated within the
$\rm R_\xi$ gauge).
After the ET has been applied,
the resulting expressions are easier to handle
because Feynman diagrams with external scalar fields have a simpler
structure and no gauge cancellations
occur when summing up the single
Feynman diagrams.

The most important fields of application of the ET are the standard
model
for the case that the Higgs boson is very heavy ($M_H\gg M_W$)
(heavy-Higgs Standard model, HHSM) and the gauged nonlinear
$\sigma$-model (GNLSM) \cite{bash,apbe},
which is the formal limit $M_H\to \infty$ of the standard model.
In these cases the self interactions
of the (physical and
unphysical) scalar fields
become strong and thus, due to the ET, the
interactions among  the longitudinal vector bosons are strong
\cite{lequth,chga,chgoge,dawi,ch,dohe,veve}. Scattering of
longitudinal vector bosons ($V_LV_L\to V_LV_L$) is the
phenomenologically most interesting process with
respect to investigations of the physics
of a strongly interacting scalar sector
in future experiments like LHC.

However it should be noted that the proofs of the ET in
\cite{chga,gokone} are only valid for the standard model with
a light Higgs boson ($E\gg
M_H$); they do not directly apply to the HHSM or the GNLSM. To
illustrate this, I briefly scetch the
essential steps of the proof:
\begin{enumerate}
\item First the BRS invariance
of the quantized Lagrangian is used in
order to derive the identity \cite{gokone}
\begin{equation}
<A|T(F_{a_1}\cdots F_{a_n})|B>=0 \label{wai}
\end{equation}
with $|A>$, $|B>$ being physical states and with the $\rm R_\xi$
gauge-fixing conditions being $F_{a_i}=0$.
This identity is a consequence of BRS invariance alone and
thus it is valid in each gauge theory, i.e.,
it holds for theories
with a light Higgs boson, with a heavy Higgs boson and without a
Higgs boson like the GNLSM; it even holds for Lagrangians with
additional effective interaction terms of higher dimension
\cite{dope,hekuli,gkku}.
\item The identity (\ref{wai}) implies the relation \cite{chga,ve}
\begin{eqnarray}&&
{\cal M}(V_{L,1}\ldots V_{L,N_1}A\to V_{L,1}\ldots V_{L,N_2}B)
\nonumber\\
&&
=\sum_{M_1=0}^{N_1}\sum_{M_2=0}^{N_2}i^{M_1}(-i)^{M_2}
[{\cal M}(\varphi_{1}\ldots\varphi_{M_1}
v_{M_1+1}\ldots v_{N_1}
A\to \varphi_{1}\ldots\varphi_{M_2}v_{M_2+1}\ldots v_{N_2}B)
\nonumber\\
&&\qquad\qquad\qquad\qquad\qquad
 +\mbox{permutations of the $\varphi$s and $v$s}],
\label{get}\end{eqnarray}
which expresses the amplitude ${\cal M}$
for a scattering process with external
longitudinal vector bosons $V_L$
($A$ and $B$ denote the other in- or outgoing particles)
as the sum of all amplitudes in
which each longitudinal vector boson is either replaced by the
corresponding pseudo-Goldstone boson $\varphi$ or its polarization
vector $\epsilon_L$ by the nonleading part
\begin{equation}
v^\mu=\epsilon_L^\mu-\frac{P^\mu}{M}.\label{v}
\end{equation}
Also (\ref{get}) is of general validity; I will call it the
generalized equivalence theorem (GET).
\item Finally, the fact that the amplitudes in the standard model
do not increase with increasing energy $E$
for $E\gg M_H$ is used in order to show that all amplitudes on the
r.h.s.\ of (\ref{get}) which have external $v$s behave
at most as $O(E^{-1})$
at high energies. The only $O(E^0)$ contribution comes from the
amplitude in which {\em all\/} $V_L$s are replaced by
$\varphi$s \cite{chga,ve}. This
is the statement of the ET in this case.
\end{enumerate}
It is obvious that the third step of the proof does not apply to
the HHSM or the GNLSM, where the $S$-matrix elements increase with
increasing energy.
Furthermore, the ET, as proven in \cite{chga,gokone}
only states that {\em external\/} longitudinal vector bosons can be
replaced by scalars, it makes no statement about {\em internal\/}
lines. In order to argue that the interactions
of longitudinal vector bosons in the HHSM or in the GNLSM
can be derived from the scalar sector of these models,
one has to show that only diagrams with all external
{\em and\/} internal lines
being scalar lines contribute to the leading part of the amplitude.

However all considerations about the validity (or
invalidity) of the ET within a specific theory
can be based on the GET (\ref{get}) \cite{ve,gkku}
because this is valid in every
gauge theory.
In this paper I will develop
a simple power-counting method with that one can easily determine
which of the diagrams corresponding to the r.h.s. of (\ref{get})
for a given process at a given loop order
contribute to the leading part of the amplitude; i.e.\ to
that part in which the sum of the powers of the total energy $E$ and
of the Higgs mass\footnote{In the Higgs-less and
nonrenormalizable GNLSM one has to count powers of $E$ and
of the cut-off
$\Lambda$.} $M_H$ is maximum.
I will apply this method to the HHSM and to the GNLSM with the result
that the leading contribution to all
$S$-matrix elements (except for
those that decrease with increasing energy)
stems from diagrams in which all
external $V_L$s are replaced
by $\varphi$s; i.e. the ET is valid in this
case. Furthermore, for processes with all external particles being
longitudinal vectors, the leading diagrams are those in which all
lines, external and internal ones, are scalar lines, i.e.\ there are
even no contributions from internal vectors, fermions or ghosts
(at an arbitrary loop order).

The latter result can easily be visualized by
an intuitive argument. (Actually, it has already frequently been used
in the literature \cite{lequth,chga,chgoge,dawi,ch,dohe,veve}.)
The scalar self interactions in the HHSM and in the GNLSM are strong
(in the the first case they are $O(M_H^2/M_W^2)$
in the second case they are $O(P_i^\mu P_j^\nu /M_W^2)$).
All diagrams on the r.h.s.\ of (\ref{get})
which have internal vector lines or external $v$s have
couplings of ordinary electroweak strength;
but those with exclusively
scalar lines have only strong couplings and thus
are the leading ones.
The power-counting method described in this article puts
this intuitive argument in a rigorous form.

This power-counting method
is an extension of the one developed by
H.~Veltman in \cite{ve} for the HHSM. The difference
between the present article and \cite{ve} is that in
\cite{ve} internal vector, fermion and ghost lines are neglected from
the beginning (which is correct in the most interesting case, as
mentioned above), while I consider first all types of
internal lines and then, based of the results of the power counting,
I exame to what extent diagrams with
internal lines other than scalar lines
can be neglected in the leading order.

Furthermore I extend the power-counting method to the GNLSM.
Although this is formally constructed as the limit $M_H\to\infty$ of
the standard model
\cite{apbe}, $S$-matrix elements calculated within
the HHSM and the GNLSM only agree at the tree level but not at higher
orders of the perturbation theory\footnote{In order to carry out
calculations beyond the tree-level in a nonrenormalizable theory
like the GNLSM, one has to introduce a
cut-off $\Lambda$. It turns out that
$S$-matrix elements
calculated in the HHSM and the GNLSM do {\em not\/}
correspond to each other after the replacement $\Lambda \to
M_H$ \cite{veve,bive}.}
\cite{veve,bive}, and thus the
latter has to be treated seperately.
It turns out, however, that the
results concerning the validity of the ET are the same in both cases.

As an example, I will apply the GET (\ref{get}) and the
power-counting method to
vector-boson scattering $VV \to VV$ with arbitrary polarization
states. As mentioned,
for all four $V$s being longitudinal the leading
contribution comes from diagrams with only (external and internal)
scalar lines.
But also in the other cases
one finds interesting
results. If there are two transversal and two
longitudinal $V$s, the ET is also valid
but there are contributions from
diagrams with internal vector lines. However
all lines in loops are
scalars. For four transversal $V$s, there are leading one-loop
contributions from diagrams with vector, fermion and ghost lines in
the loop, but at higher loop orders again diagrams with only scalar
lines in the loops are dominating.

I also apply the power-counting method to the standard model
with a light Higgs boson in order to point out the differences to the
heavy-Higgs case and to give another illustration of this method.
This yields an alternative derivation of the ET for this case.

Finally, I apply this method to effective Lagrangians
with additional anomalous
interaction terms of higher dimension. At the tree
level the validity of the ET within such models has been examined in
\cite{gkku}, here I extend the results of \cite{gkku}
to arbitrary loop orders.
It turns out that for effective Lagrangians which contain strong
couplings among the scalar fields the ET is valid, however, that for
Lagrangians with strong couplings between the scalar
and vector fields or among the
vector fields this theorem fails. Some specific examples will be
discussed. However in any case the calculations become simplified by
applying the GET (\ref{get}) and the power-counting method.

This article is organized as follows: In Section~2 the power-counting
method based on the generalized equivalence theorem (\ref{get})
is described within the heavy-Higgs standard model. A
simple formula is derived
with which the leading diagrams for a given process
at an arbitrary loop order can be determined.
In Section~3 the
same is done within the gauged nonlinear $\sigma$-model.
In Section~4 this formula is applied to vector boson scattering.
The validity of the equivalence theorem
is shown.
In Section~5 the power-counting method is applied to the light-Higgs
standard model and
in Section~6 to effective Lagrangians. The validity
or invalidity of the equivalence theorem within several effective
theories is discussed. Section~7 contains the summary.

\section{The Heavy-Higgs Standard model}
\typeout{Section 2}
\setcounter{equation}{0}
In this section I consider the heavy-Higgs standard model at high
energies, i.e. the case
\begin{equation}
M_H, E\gg M_W, M_{f,i}, \label{hhsm}\end{equation}
where $E$, $M_H$,
$M_W$ and  $M_{f,i}$ are the total energy, the
Higgs mass, the gauge boson mass and the fermion masses
respectively.
The leading parts of the $L$-loop amplitude for a given process
are those terms in which $N$, defined as
\begin{equation}
N=\mbox{power of $M_H$}+\mbox{power of $E$}, \label{N}
\end{equation}
becomes maximum. In order to calculate this leading contribution
with help of the GET (\ref{get}) one has to proceed as
follows:
\begin{enumerate}
\item Apply the GET (\ref{get}), which expresses
the $S$-matrix element  for the process
as the sum of all amplitudes in which each external
$V_L$ is either
replaced by an unphysical Goldstone boson or its polarization
vector $\epsilon_L$ by the nonleading part $v$ (\ref{v}).
\item Construct all
Feynman diagrams that correspond to these amplitudes at $L$-loop
order (in the $\rm R_\xi$ gauge).
\item
Determine for each diagram the maximum $N$ (\ref{N})
to which it contributes.
\item Determine the leading diagrams, viz. those
for which $N$ is maximum.
\item Calculate these Feynman diagrams.
\end{enumerate}
In this section I will dicuss item 3. I will apply power counting in
order to derive a simple formula for the $N$ of the leading
contribution of a given diagram. As mentioned, this procedure is an
extension of the power-counting method developed in \cite{ve}. The
determination of the leading diagrams (item 4) will be discussed in
section~4.

Consider one single Feynman diagram at $L$-loop order with $E$
external lines, $I$ internal lines and $V$ vertices.
$E_i$, $I_i$ or $V_i$ denote the number of lines or vertices of a
specific type, where a subsript
$\phi$ stands for a scalar (both a Higgs and a
pseudo-Goldstone scalar) and $V$, $f$ and $g$
stand for a vector, an (anti)fermion
and an (anti)ghost, respectively. (E.g.\
$I_f$ is the number of internal
fermion lines and $V_{VV\phi}$ is the
number of vector-vector-Higgs and
of vector-vector-Goldstone vertices.) $E_v$ is the number of external
longitudinal vectors whose polarization vectors are replaced by $v$s
(\ref{v}).
Furthermore I define
\begin{eqnarray}
V_\phi&=&V_{\phi\phi\phi}+V_{\phi\phi\phi\phi}\nonumber\\
V_d&=&V_{V\phi\phi}+V_{VVV}+V_{Vgg},\nonumber\\
V_f&=&V_{ff\phi}+V_{ffV},\nonumber\\
V_{0}&=&V-V_\phi-V_d-V_f, \label{v2}\end{eqnarray}
which denote the numbers of
scalar self-couplings, derivative couplings, fermionic
couplings and of the remaining vertices
respectively, and
\begin{equation} I_0=I-I_f=I_V+I_\phi+I_g.\end{equation}

Counting powers of $E$ from external lines and powers of $M_H$ from
the vertex factors one finds for the contribution ${\cal M}_i$
of this diagram to the $S$-matrix element
\begin{equation}
{\cal M}_i=c E^{\frac{1}{2}E_f-E_v} M_H^{2V_\phi} I_F,
\label{M}\end{equation}
where $c$ is a constant and $I_F$ is the part of the
the Feynman intergral
\begin{equation}
\int d^4P_1\cdots d^4P_L \frac{P_1 \cdots P_{V_d}}
{(P_1^2-M_1^2)\cdots (P_{I_0}^2-M_{I_0}^2)
(P\mbox{\settowidth{\zeichen}{$\displaystyle P $}%
\hspace{-0.5\zeichen}%
\hspace{-0.5\strich}}/_{I_0+1}-M_{I_0+1})\cdots
(P\mbox{\settowidth{\zeichen}{$\displaystyle P $}%
\hspace{-0.5\zeichen}%
\hspace{-0.5\strich}}/_I-M_I)}\label{if}\end{equation}
that remains after renormalization.
(Lorentz indices are not explicitly written here because they are
unimportant for the power counting.) The $P_i$ are the internal
particles' momenta and the $M_i$ are their masses. The leading part
of $I_F$ has the form
\begin{equation}
I_F= a_0 M_H^D+ a_1 M_H^{D-2} E^2+
a_2 M_H^{D-4}E^4+
\cdots \label{iflead}\end{equation}
where $D$ is the dimension of $I_F$. The coefficients $a_i$ may
depend on logarithms of $M_H$ or of $E$ but not on powers.
Due to the screening theorem
some of them may be zero or cancel
when summing the contributions of the single
diagrams \cite{dawi,bive,mve} .
However this does not affect the power counting
since I count powers of $M_H$ and of $E$ simultaneously
and thus each term in (\ref{iflead}) contributes to the same $N$
(\ref{N}). The dimension of $I_F$ can be read from (\ref{if})
\begin{eqnarray} D&=&4L+V_d-2I_0-I_f\nonumber\\
&=&4L+V_d-2I+I_f.\end{eqnarray}
This, together with (\ref{M}) and (\ref{iflead}) yields
\begin{eqnarray} N&=&D+2V_\phi+\frac{1}{2}E_f- E_v\nonumber\\
&=& 4L+2V_\phi+V_d-2I+I_f+\frac{1}{2}E_f- E_v.\label{zw1}
\end{eqnarray}
With
\begin{equation} L=I-V+1 \label{L}\end{equation}
one obtains
\begin{eqnarray}
N&=&2L+2-2V+2V_\phi+V_d+I_f+\frac{1}{2}E_f-E_v\nonumber\\
&=&2L+2-V_d-2V_0-2V_f+I_f+\frac{1}{2}E_f-E_v.\end{eqnarray}
Using
\begin{equation}
\frac{1}{2}E_f+I_f=V_f\end{equation}
one finally finds
\begin{equation}
N=2L+2-M\qquad \mbox{with} \qquad M=V_d+2V_0+V_f+E_v.
\label{NM}\end{equation}
This formula gives the
$N$ (\ref{N}) of the leading part of
each diagram contributing at
$L$-loop order to the r.h.s. of (\ref{get}) as a function of the
number of external $v$s and of the number of the different types of
vertices. In
Section~4 I will demonstrate how the leading diagrams for
a given process can easily be determined from (\ref{NM}).

\section{The Gauged Nonlinear $\bf \sigma$-Model}
\typeout{Section 3}
\setcounter{equation}{0}
In this section I proceed analogously to the last one, however I
consider the gauged nonlinear $\sigma$-model.
In this model the spontaneous
breakdown of the gauge symmetry is nonlinealy realized, such that no
physical Higgs boson exists. There are only three unphysical
pseudo-Goldstone bosons and
nonpolynomial interactions
of these with each other and with the vector
bosons, fermions and ghosts \cite{bash,apbe}.
This model represents an alternative way
to parametrize a strongly interacting
scalar-sector.

Since the GNLSM is nonrenormalizable, one has to introduce a cut-off
$\Lambda$ so that $S$-matrix elements calculated beyond the
tree-approximation become finite\footnote{Usually one applies
dimensional regularization in order to calculate Feynman integrals
and then identifies terms with poles at space-time dimension $D=4$
as being logarithmically cut-off dependent, terms with poles at
$D=2$ as being
quadratically cut-off dependent, etc.\ \cite{apbe,bive}.}
\cite{apbe}. Thus
the GNLSM has to be interpreted as an effective theory
which is the low-energy approximation of unknown new physics at the
TEV scale. Usually the cut-off $\Lambda$ is assumed to be the
scale of the new physics.

I consider the GNLSM at high energies, i.e.\ the case
\begin{equation} \Lambda > E \gg M_W, M_{f,i}. \label{gnlsm}
\end{equation}
The leading
diagrams in this case are those for which
\begin{equation}
N=\mbox{power of $\Lambda$}+\mbox{power of $E$}, \label{NSM}
\end{equation}
is maximum.

Because of the nonpolynomial structure of the GNLSM Lagrangian, there
are vertices with an arbitrary number of legs. In addition to the
Yang-Mills, vector-fermion and vector-ghost couplings there are
$VV\phi^N$
($N\ge 1$), $ff\phi^N$  ($N\ge 1$) and $gg\phi^N$ ($N\ge 1$)
couplings without derivatives, $V\phi^N$ ($N\ge 2$) couplings
with one
derivative and $\phi^{2N}$ ($N\ge 2$) couplings  with two
derivatives. It turns out that in spite of these nonpolynomial
couplings the power-counting method of the last section can easily
be applied to the GNLSM.

I consider again a Feynman diagram at $L$-loop order and define,
in analogy to (\ref{v2}),
\begin{eqnarray}
V_{\phi}&=&\sum_{N=2}^\infty V_{\phi^{2N}},\nonumber\\
V_d&=&\sum_{N=2}^\infty V_{V\phi^N}+V_{VVV}+V_{Vgg},\nonumber\\
V_f&=&\sum_{N=1}^\infty V_{ff\phi^N}+V_{ffV},\nonumber\\
V_0&=&V-V_\phi-V_d-V_f.\label{vnl}\end{eqnarray}
The contribution of the diagram
to the amplitude has the form
\begin{equation} {\cal M}_i=c E^{\frac{1}{2}E_f-E_v} I_F,
\label{M2}\end{equation}
the Feynman intergral is
\begin{equation}
\int d^4P_1\cdots d^4P_L \frac{P_1 \cdots P_{V_d}P_{V_d+1}
\cdots P_{V_d+2V_\phi}}
{(P_1^2-M_1^2)\cdots (P_{I_0}^2-M_{I_0}^2)
(P\mbox{\settowidth{\zeichen}{$\displaystyle P $}%
\hspace{-0.5\zeichen}%
\hspace{-0.5\strich}}/_{I_0+1}-M_{I_0+1})\cdots
(P\mbox{\settowidth{\zeichen}{$\displaystyle P $}%
\hspace{-0.5\zeichen}%
\hspace{-0.5\strich}}/_I-M_I)}\label{ifnl}\end{equation}
and the leading part of
$I_F$ can be written as
\begin{equation} I_F= a_0 \Lambda^D+ a_1 \Lambda^{D-2} E^2+
a_2\Lambda^{D-4}E^4+
\cdots \label{ifleadnl}\end{equation}
(The coefficients $a_i$ are in general different from
those found in the HHSM (\ref{iflead}) \cite{veve,bive}.)
The dimension $D$ of $I_F$ follows from (\ref{ifnl})
\begin{eqnarray} D&=&4L+2V_\phi+V_d-2I_0-I_f\nonumber\\
&=&4L+2V_\phi+V_d-2I+I_f.\end{eqnarray}
Thus one finds
\begin{eqnarray} N&=&D+\frac{1}{2}E_f-E_v\nonumber\\
&=& 4L+2V_\phi+V_d-2I+I_f+\frac{1}{2}E_f- E_v,\label{zw2}
\end{eqnarray}
which is identical with (\ref{zw1}). Now one can proceed as in the
previous section and obtains (\ref{NM}) as the final result.

Thus, although the vertices in the GNLSM have a different
structure than those
in the HHSM, one finally ends up with the same formula
for $N$. The only difference is that $N$ is defined as (\ref{NSM})
instead of (\ref{N}). Because of this result the determination of the
leading diagrams and the derivation of the ET can now be done
simultaneously for the HHSM and the GNLSM.

\section{Determination of the Leading Diagrams}
\setcounter{equation}{0}
\typeout{Section 4}
Now I discuss item 4 of the program outlined at the begin of
Section~2. Having constructed all Feynman diagrams that correspond
at $L$-loop order to the r.h.s.\ of (\ref{get}) for a given process,
one has to determine those diagrams, for which $N$ (\ref{NM}) is
maximum, i.e. those for which $M$ is minimum. In order to obtain the
leading contributions it is then sufficient to calculate only these
diagrams. As mentioned, this discussion applies to both the HHSM and
the GNLSM.

For processes with all external particles being longitudinal vector
bosons, e.g.\ $V_LV_L\to V_LV_L$,
those diagrams contributing to the r.h.s.\ of (\ref{get})
in which all (external and internal) lines are scalar lines have
\begin{equation} V_d=V_0=V_f=E_v=0\label{llll1}\end{equation}
and thus
\begin{equation} M=0, \qquad N=2L+2, \label{llll2}\end{equation}
while for diagrams with
external $v$s or with internal vector, fermion or ghost lines
one finds $M>0$, $N<2L+2$; i.e. these are nonleading.
This means, for these processes the ET is valid and, even more, the
interactions of the
longitudinally polarized vector bosons in the HHSM
or in the GNLSM follow from the scalar self-interactions in these
models.

With help of Eq.\ (\ref{NM}) one can also determine the leading
contribution to processes with external transversal vector bosons
$V_T$ or fermions, although
then diagrams which fulfil (\ref{llll1}) cannot be constructed.
As an example let me discuss the process $VV\to VV$ with all possible
polarizations of the $V$s. (The theoretically most interesting
contribution comes of course from $V_LV_L \to V_LV_L$ but
vector-boson scattering with some or all in- and outgoing $V$s being
transversal yields a background to this process and thus is
phenomenologically important, too.)
\begin{itemize}
\item If there is one external $V_T$ and three $V_L$s,
the $V_T$ can be coupled
by one derivative coupling to the scalar field. These diagrams have
$V_d=1, \; V_0=V_f=E_v=0$ and thus
\begin{equation} M=1, \qquad N=2L+1.\end{equation}
All other diagrams have a smaller $N$. I.e.,
like in the case of four longitudinal $V$s, only
diagrams with all external $V_L$s
being replaced by $\varphi$s and with only
internal scalar lines contribute in the leading order.
\item If there are two longitudinal and two transversal $V$s, in the
leading diagrams each of the $V_T$s
is either seperately coupled by a derivative
coupling to the scalar fields
($V_d=2, \; V_0=V_f=E_v=0$, e.g.\ Figure~1a)
or both are coupled by one vertex to the scalar fields ($V_0=1,\;
V_d=V_f=E_v=0$, e.g.\ Figure~1b)
or by one vertex to an internal vector line, which
couples by a derivative coupling to the scalar fields
($V_d=2, \; V_0=V_f=E_v=0$, e.g. Figure~1c).
In all cases one finds
\begin{equation} M=2, \qquad N=2L.\end{equation}
Diagrams with external $v$s are nonleading but there are
leading diagrams with
internal vector lines at an
arbitrary loop order (Figures~1c and 1d).
However there are no leading diagrams with internal $V$ lines {\em in
loops\/}.
\item If there are three transversal and one longitudinal
external $V$, one finds by a similar argumentation that the leading
diagrams have
\begin{equation} M=3, \qquad N=2L-1\end{equation}
with either $V_d=3,\; V_0=V_f=E_v=0$ or $V_0=1, \; V_d=1,\;
V_f=E_v=0$. There are leading contributions from
diagrams with internal $V$s but without
$V$-lines in loops. At the tree-level
there are in addition
leading diagrams in which the $V_L$
is replaced by a $v$ ($E_v=1,\; V_d=2,\; V_0=V_f=0$,
Figure~2a, or
$E_v=1,\; V_0=1,\; V_d=V_f=0$, Figure~2b) but for $L\ge 1$ the $V_L$
has always to be replaced by a $\varphi$ in the leading order.
\item For $V_TV_T \to V_TV_T$ one finds for $L\ge 1$
\begin{equation} M=4, \qquad N=2L-2.\end{equation}
(At the tree level ($L=0$) one has $M=2, \; N=0$ with the leading
diagrams corresponding to Figure~2 with a $V_T$ instead of
the $v$.) At the
one-loop level there are leading diagrams with internal
vector, fermion or ghost lines
even in the loop (Figure~3a) but for
$L\ge 2$ internal $V$ lines may only occur outside the loop; in the
loops there are
only scalar lines (Figure~3b). I.e.\ even for this process
without external longitudinal vectors, calculations become simplified
because of the strongly interacting scalar sector at two and
higher loop orders.
\end{itemize}
In summary one finds that for all amplitudes with $N>0$ in
the leading diagrams  all $V_L$s are replaced by $\varphi$s;
i.e.\ the ET is valid.

As an example for a process with external fermions, I consider
$f\bar{f}\to V_LV_L$. There are diagrams with $V_f=1,\;
V_d=V_0=E_v=0$ and thus $M=1,\; N=2L+1$, however these have a
scalar-fermion coupling and thus are suppressed if $f$ is a light
fermion. The leading diagrams in this case have one internal vector
line coupled to the fermion fields and to the scalar fields
but no vector or fermion lines in
loops ($V_d=V_f=1,\;  V_0=E_v=0,\; M=2,\; N=2L$).

Finally I want to discuss the following points:
\begin{itemize}
\item Before doing power counting one neccesarily
has to apply the GET (\ref{get})
because otherwise
cancellations take place when summing up
the single diagrams with external longitudinal vector bosons.
After (\ref{get}) has been applied,
all external $V_L$s are replaced by
$\varphi$s or by $v$s and no more cancellations occur.
For instance, for $V_L V_L \to V_L V_L$ the single
tree-level diagrams with
external $V_L$s have $N=4$
but the resulting amplitude has only $N=2$. However,
the diagrams contributing to the r.h.s of (\ref{get}) for $L=0$
have at most $N=2$ seperately (see above).
\item $N$ (\ref{N})
denotes the {\em sum} of the powers of $E$ and of $M_H$
because both $E$ and $M_H$ are assumed to be large in comparison
to $M_W$ and the $M_{f,i}$. The results of this section
apply thus to both cases, $M_H\ge E$ and $E\ge M_H$ as long as
(\ref{hhsm}) is fulfilled.
\item One could be tempted to count only powers of $M_H$
(or of $\Lambda$ in the GNLSM)
because the largest powers of $M_H$ are
dominating for $M_H> E\gg M_W,
M_{f,i}$. However, the leading power
of $M_H$ {\em alone\/}
cannot be determined by power counting, because, due to the
screening theorem \cite{dawi,bive,mve},
this cancels or becomes absorbed by
renormalization. For example, by power counting one would expect an
$M_H^2$ dependence of the $V_LV_L \to V_LV_L$ amplitude at the
tree level and an $M_H^4$ dependence at the one-loop level, but
the amplitude behaves as $M_H^0$ for $L=0$ and as
$\ln(M_H)$ for $L=1$ \cite{dawi}. However simultaneous counting of
powers of $E$ and of $M_H$ yields $N=2L+2$ (\ref{llll2})
for this case
and actually the leading terms are proportional
to $E^2$ at the tree level and to $E^4$ at the one loop level.
Thus, this power counting method can only be used
in order to determine the
diagrams which are leading in $M_H$ (or $\Lambda$)
{\em and} in $E$ but not to
disentangle contributions with large powers of $M_H$ from those with
large powers of $E$. It is obvious from the above procedure that
this is sufficient to derive the equivalence theorem for the HHSM
or the GNLSM.
\item Eq.\ (\ref{NM}) may also be used in order to determine the
next-to-leading contributions to a given process, if one is
interested in these.
\item The leading part of an $S$-matrix element has the structure
\begin{equation}  c_0 M_H^N+c_1 M_H^{N-2}E^2+c_2 M_H^{N-4}E^4+\ldots
\label{lc}\end{equation}
(in the GNLSM one has to replace $M_H$ by $\Lambda$), while the
next-to leading part has the form
\begin{equation}
\sum_i m_i^2(d_{0,i} M_H^{N-2}+ d_{1,i} M_H^{N-4} E^2+\cdots)
\label{ntlc}\end{equation}
with $m_i$ being $M_W$ or $M_{f,i}$.
Due to the screening theorem some of the $c_n$ in (\ref{lc}) may
vanish. To ensure that nevertheless the contribution of (\ref{lc}) is
much larger than that of (\ref{ntlc}), in additon to (\ref{hhsm})
the condition
\begin{equation} \frac{M_H}{E}\ll\frac{E}{M_W},\frac{E}{M_{f,i}}
\label{ec1}\end{equation}
(for $M_H\ge E$ and for the GNLSM with $M_H\to \Lambda$)
or
\begin{equation} \frac{E}{M_H}\ll\frac{M_H}{M_W},\frac{M_H}{M_{f,i}}
\label{ec2}\end{equation}
(for $E\ge M_H$) has to be fulfilled.
This is a realistic assumption for investigations of vector-boson
scattering within the strongly-interacting-Higgs
scenario \cite{lequth,chga,chgoge,dawi,ch,dohe,veve},
where one assumes
that the total energy $E$ is already close to the scale of
the new physics which corresponds to $\Lambda$ or to $M_H$.
\item If one applies the GET (\ref{get}) beyond the tree level,
correction factors stemming from the renormalization of external
lines have to be considered for each $V_L$ which is replaced by a
$\varphi$ \cite{dope,hekuli,yayu}. The correct form of (\ref{get}) is
thus
\begin{eqnarray}\!\!\!\!\!\!\!\!\!\!\!\!&&
{\cal M}(V_{L,1}\ldots V_{L,N_1}A\to V_{L,1}\ldots V_{L,N_2}B)
\nonumber\\\!\!\!\!\!\!\!\!\!\!\!\!
&&
=\sum_{M_1=0}^{N_1}\sum_{M_2=0}^{N_2}(iC)^{M_1}(-iC)^{M_2}
[{\cal M}(\varphi_{1}\ldots\varphi_{M_1}
v_{M_1+1}\ldots v_{N_1}
A\to \varphi_{1}\ldots
\varphi_{M_2}v_{M_2+1}\ldots v_{N_2}B)
\nonumber\\\!\!\!\!\!\!\!\!\!\!\!\!
&&\qquad\qquad\qquad\qquad\qquad
 +\mbox{permutations of the $\varphi$s and $v$s}],
\label{getc}\end{eqnarray}
with the $C$s being the correction factors. However,
it has been shown that a
renormalization scheme exits in which the $C$s
(calculated at one-loop order) do not depend on $M_H$ \cite{yayu}
and thus they do not contribute to $N$ (\ref{N}). Therefore,
to consider these
correction factors
to an $L$-loop diagram means to consider additional
loop-corrections to the external lines, i.e. contributions
from diagrams with a
loop order higher than $L$ but with the same $N$ (\ref{N}).
Since in the HHSM and the GNLSM
the $N$ of the leading contribution increases with increasing $L$,
these corrections are nonleading effects\footnote{Although
higher loop contributions to the $C$s have
not been calculated, one should expect that,
due to the screening theorem, these also depend on smaller powers
of $M_H$ as would be necessary in order to be leading effects.}.
Thus, in order to calculate
the leading terms, the $C$s in (\ref{getc}) can be
neglected\footnote{The
situation is different in the light-Higgs standard model,
where $N$ (defined as the power of $E$ only in that case)
does not grow with $L$.} if
an adequate renormalization scheme is applied.
\end{itemize}

\section{The Light-Higgs Standard Model}
\setcounter{equation}{0}
\typeout{Section 5}
This power-counting method can also be applied to the light-Higgs
standard model (LHSM) at high energies, i.e. the case
\begin{equation} E\gg M_H, M_W, M_{f,i}. \label{lhsm}\end{equation}
Here one has to count only powers of $E$, i.e.
the leading diagrams are those for which $N$, defined as
\begin{equation} N=\mbox{Power of $E$}, \end{equation}
is maximum.
One can now proceed in analogy to
the discussion of the HHSM in Section~2.
In the LHSM the factor $M_H^{2V_\phi}$
in (\ref{M}) does
not contribute to $N$. Thus in (\ref{NM}) one has to
subtract $2V_\phi$ and finds
\begin{equation} N=2L+2-2V_\phi-V_d-2V_0-V_f-E_v. \end{equation}
Using (\ref{L}) and
\begin{equation} E+2I=3V_3+4V_4 ,\end{equation}
with $V_3$ and $V_4$ being the numbers of vertices with 3 and 4 legs,
respectively,
\begin{eqnarray}
V_3&=&V_d+V_f+V_{\phi\phi\phi}+V_{VV\phi}+V_{gg\phi},\nonumber\\
V_4&=&V_\phi+V_0-V_{\phi\phi\phi}
-V_{VV\phi}-V_{gg\phi},\end{eqnarray}
one finds
\begin{equation}
N=4-E-\tilde{M} \qquad \mbox{with} \qquad \tilde{M}=
V_{\phi\phi\phi}+V_{VV\phi}+V_{gg\phi}+E_v.
\label{NM2}\end{equation}

The leading diagrams for a given process are thus those without
$\phi\phi\phi$-,  $VV\phi$- or $gg\phi$-vertices and without external
$v$ lines (if it is possible to construct such diagrams).
These have
\begin{equation}
\tilde{M}=0, \qquad N=4-E
\label{smlead}\end{equation}
at an arbitrary loop-order $L$, i.e.\ $N$ does not grow with
increasing $L$ unlike in the HHSM.
This result shows the validity of the
ET in this case and the perturbative
unitarity of the
LHSM. However, in distinction from the HHSM, there are leading
contributions from diagrams with all types of internal lines, i.e.
in the LHSM the ET is indeed only a statement about the external
lines.

For some processes it is not possible to construct diagrams which
fulfil (\ref{smlead}).
For instance for $VV\to VV$ with one transversal
and three longitudinal $V$s the
leading diagrams (at each loop order) contain
at least either a
$\phi\phi\phi$-,  $VV\phi$- or $gg\phi$-vertex (Figure~4a)
or an external $v$ line (Figure 4b). These have
$\tilde{M}=1, \; N=-1$. This is not
a contradiction to the ET, which is
only a statement about the $O(E^0)$ ($N=0$) contributions
\cite{chga,gokone,ve}.

\section{Effective Lagrangians}
\setcounter{equation}{0}
\typeout{Section 6}
The power counting method
described above can also be used in order to
simplify the calculation of $S$-matrix elements within
effective theories with Lagrangians of the
type
\begin{equation} {\cal L}_{eff}={\cal L}_0+\sum_i
\epsilon_i{\cal L}_i,\label{leff}\end{equation}
where ${\cal L}_0$ is either
the standard-model Lagrangian (with a light or a
heavy Higgs boson) or the GNLSM Lagrangian and the ${\cal L}_i$ are
nonstandard interaction terms of higher dimension. These are gauge
invariant, too,
with the scalar sector being linearly realized in the first
case and nonlinearly realized in the second case \cite{apbe,llr}.

Since effective theories are nonrenormalizable, $S$-matrix elements
calculated in higher loop-orders depend on a cut-off $\Lambda$. The
leading diagrams in the no-Higgs or light-Higgs scenario,
are thus those for which $N$, defined as (\ref{NSM}), is maximum.
In the heavy-Higgs scenario $N$ has to be defined as the sum of the
powers of $E$, $M_H$ and $\Lambda$.

In this section I will discuss
nonlinear effective Lagrangians
without a Higgs boson (chiral Lagrangians)
but the discussion can easily be extended to
effective Lagrangians with a heavy Higgs boson (for which one finds
analogous results) or with a light Higgs boson (for which one finds
different results). Furthermore I restrict to Lagrangians which
contain only effective interactions of the vector and scalar fields
but no anomalous fermionic interactions.

Consider again a given Feynman diagram.
Let $V_i^\epsilon$ be the number of $i$-derivative vertices stemming
from the nonstandard terms (proportional to $\epsilon_i$) and
\begin{equation} V^\epsilon=\sum_iV_i^\epsilon\end{equation}
the number of all nonstandard vertices in this diagram.
In analogy to the procedure in Section~3 one finds
\begin{equation} N=2L+2-M\qquad \mbox{with} \qquad
M=V_d+2V_0+V_f+E_v+\sum_i(2-i)V_i^\epsilon
\label{NMeff}\end{equation}
where $V_d$, $V_0$ and $V_f$ denote the number of {\em standard\/}
vertices of a specific type only.
(One can see that in this  case $M$ may become
negative and thus $N>2L+2$.) With this formula the leading
diagrams for fixed $L$ and $V^\epsilon$ within a given effective
theory for a given process can be determined.

To give some examples
I discuss the leading contributions of some
specific effective interaction
terms to the process $V_L V_L \to V_L V_L$.
I use the follwing notation:
\begin{eqnarray}
W_\mu&=&\frac{1}{2}W_{\mu i}\tau_i,\nonumber\\
W_{\mu\nu}&=&\partial_\mu W_\nu-\partial_\nu W_\mu+ig[W_\mu,
W_\nu],\nonumber\\
B_{\mu\nu}&=&\partial_\mu B_\nu-\partial_\nu B_\mu,\nonumber\\
U&=&\exp\left(\frac{i\varphi_i\tau_i}{v}\right),\nonumber\\
D_\mu U&=&\partial_\mu U+igW_\mu U-\frac{i}{2}
g^\prime U\tau_3B_\mu,\label{not}
\end{eqnarray}
where
$W_{\mu i}$ and $B_\mu$ denote the gauge fields, $g$ and $g^\prime$
the coupling contants, $v$ the vacuum expectation value,
$\varphi_i$ the pseudo-Goldstone fields and
$\tau_i$ the Pauli-matrices.
Since the effective couplings are assumed to be small in comparison
to the standard couplings, I consider only diagrams with one
nonstandard vertex
($V^\epsilon=1$); however this
power-counting method applies to arbitrary
$V^\epsilon$.
\begin{itemize}
\item
The effective terms
\begin{eqnarray}
\mbox{${\cal L}_{DD1}^\prime$}&
=&-\frac{1}{16}\,\mbox{tr}\,
[(D_\mu U)^\dagger(D_\nu U)]
\,\mbox{tr}\,[(D^\mu U)^\dagger(D^\nu U)],\nonumber\\
\mbox{${\cal L}_{DD2}^\prime$}&
=&-\frac{1}{16}
\,\mbox{tr}\,[(D_\mu U)^\dagger(D^\mu U)]
\,\mbox{tr}\,[(D_\nu U)^\dagger(D^\nu U)]\label{ldd}
\end{eqnarray}
contain scalar self-interactions (of at least four scalars)
with four derivatives, vector-scalar interactions with less than four
derivatives and four-vector couplings without derivatives.
Their leading contributions (for $V^\epsilon=1$) at an
arbitrary loop order
stem from diagrams
with only scalars as external or internal lines and one
nonstandard scalar self-coupling ($V_4^\epsilon=1,
\; V_d=V_0=V_f=E_v=0$).
These have $M=-2,\; N=2L+4$. I.e., the ET is valid in this model and,
even more, the the interactions of the longitudinal vector bosons
correspond to the scalar self interactions in the effective theory.
This result has already been used in \cite{dohe}.
\item
The terms
\begin{eqnarray}
{\cal L}_{W\phi}&=&\frac{1}{2}i
\,\mbox{tr}\,[(D_\mu U)^\dagger W^{\mu\nu}
(D_\nu U)],\nonumber\\
{\cal L}_{B\phi}&=&-\frac{1}{4}i
\,\mbox{tr}\,[\tau_3(D_\mu U)^\dagger
(D_\nu U)]B^{\mu\nu}\label{lwp}\end{eqnarray}
contain vector-scalar couplings
and vector
self-interactions. The couplings of one or two vectors
to the scalars contain three or two derivatives respectively.
The leading diagrams are thus those with
one vector line which couples by a three-derivative vertex to the
scalar fields. This may either be an internal line
($V_3^\epsilon=1,
\;V_d=1,\; V_0=V_f=E_v=0$, Figure~5a) or an external $v$
($V_3^\epsilon=1,\; E_v=1,\; V_d=V_0=V_f=0$, Figure~5b).
In addition there are diagrams with
$V_2^\epsilon=1,\; V_d=V_0=V_f=E_v=0$.
All these diagrams have $M=0,\; N=2L+2$. I.e., the ET cannot be
applied in order to  determine the leading contributions of these
terms because there are leading diagrams with external $v$s.
Furthermore there are contributions from
internal vector lines even in loops
(Figure~5a).
\item
The term
\begin{equation}{\cal L}_W=-\frac{2}{3\Lambda^2}i\,\mbox{tr}\,
(W_\mu^{\,\,\,\nu}
W_\nu^{\,\,\,\lambda}W_\lambda^{\,\,\,\mu})\label{lw}\end{equation}
(quadrupole term) contains
vector self-interactions. The three-vector
vertices depend on three derivatives.
The leading
contributions come from diagrams with
$V^\epsilon_3=1,\; V_d+2V_0+E_v=3,\;
V_f=0$ (Figure~6). These have $M=2,\; N=2L$.
Also in this case the ET is not valid; there
are leading diagrams with external $v$s and with internal vector
lines even in loops.
\end{itemize}
One can see that the for the effective terms ${\cal L}_{DD1}$ and
${\cal L}_{DD2}$, which contain
strong scalar self-interactions, the ET is
valid, but it is not valid for the terms ${\cal L}_{W\phi}$ and
${\cal L}_{B\phi}$
which contain strong vector-scalar interactions and
for ${\cal L}_W$ which contains strong vector self-interactions.

It should be noted that the contributions of the
terms ${\cal L}_{W\phi}$, ${\cal L}_{B\phi}$
and ${\cal L}_{W}$ to $V_L V_L \to
V_L V_L$ have the same or a smaller $N$ than those of the GNLSM and
thus only yield small corrections the GNLSM amplitude in this
process. Actually, they yield larger deviations in vector-boson
scattering with two or four external $V_T$s \cite{gkku,ku}, which can
also be discussed on the basis of eq.~(\ref{NMeff}). A complete
determination of the diagrams that yield the leading tree-level
contributions of the above and
several other effective interaction terms
to the processes $VV\to VV$ and $f\bar{f}\to VV$ with all possible
polarizations of the $V$s is given in \cite{gkku}.

\section{Summary}
\typeout{Section 7}
In this article I have shown the validity of the equivalence
theorem (which
was originally proven for the light-Higgs standard model)
for the case of
the standard model with a heavy Higgs boson and for the
gauged nonlinear
$\sigma$-model, which contains no physical Higgs boson.
The leading part of the
$S$-matrix element for a scattering process
with in- or outgoing longitudinally polarized vector bosons at
high energies
can be found by replacing the external longitudinal vector
bosons by
the corresponding pseudo-Goldstone bosons. Even more, for processes
with all external particles being longitudinal vectors, the leading
diagrams are those in which also all
{\em internal\/} lines are scalar lines.
This means that the strong interactions among the scalar fields
in the heavy-Higgs standard model and
in the gauged nonlinear $\sigma$-model
manifest themselves in the interactions among the longitudinally
polarized vector bosons.

This result can easily be made plausible by an intuitive
argument and was
thus already frequently used in the literature. Here it has been
rigorously derived on the
basis of the generalized equivalence theorem
(\ref{get}) and the power-counting method originating from \cite{ve}.
The result of this power-counting method is a simple formula
which determines the sum
of the powers of the total energy $E$, of the
Higgs mass $M_H$ (in models
with a heavy Higgs boson) and of the cut-off
$\Lambda$ (in nonrenormalizable theories) for the leading part
of the amplitude
of each Feynman diagram contributing to the $S$-matrix element of
a given process after the generalized equivalence theorem (\ref{get})
has been applied. With this formula one can easily determine the
leading diagrams, i.e. those for whose contribution the above
sum is maximum,
for any scattering process at an arbitrary loop-order.
The generalized equivalence
theorem (\ref{get}) and the power counting method
can be applied to every gauge theory, even to effective Lagrangians
with additional
anomalous interaction terms of higher
dimension, for which the equivalence
theorem is in general not valid.

\newpage

\section*{Figure Captions}
\typeout{Figure Captions}
\begin{description}
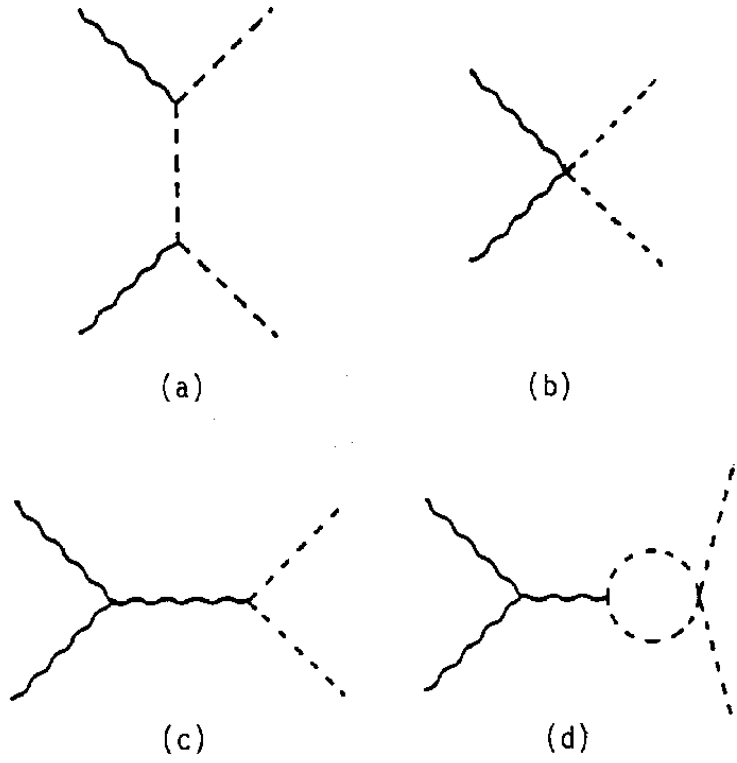
\item[Figure 1:]Some of the leading diagrams for $VV\to VV$ with two
longitudinal and two transversal $V$s at tree
(a, b, c) and one-loop level (d).
The dashed lines denote scalars, the wavy lines vectors.
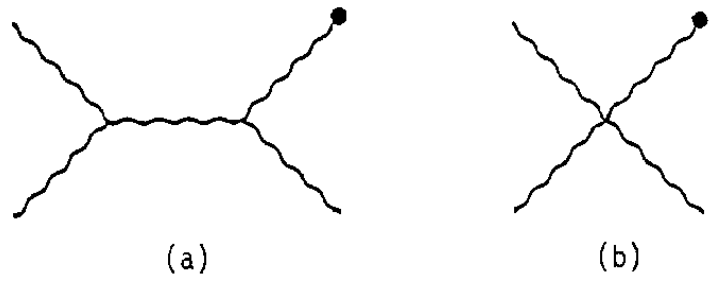
\item[Figure 2:]Some of the
leading tree-level diagrams for $VV\to VV$ with one external
$V$ being longitudinal and three being
transversal. The wavy line with a dot stands for an external $v$
line (\ref{v}).
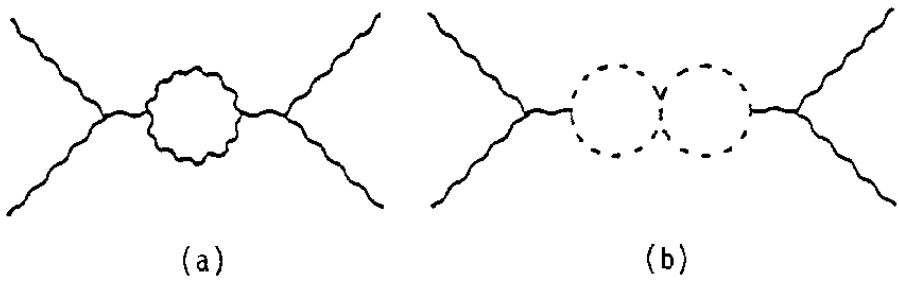
\item[Figure 3:]Some leading
diagrams for $V_TV_T \to V_TV_T$ at one- (a)
and two-loop level (b). Instead of the vector loop in (a) there may
also be a fermion or ghost loop.
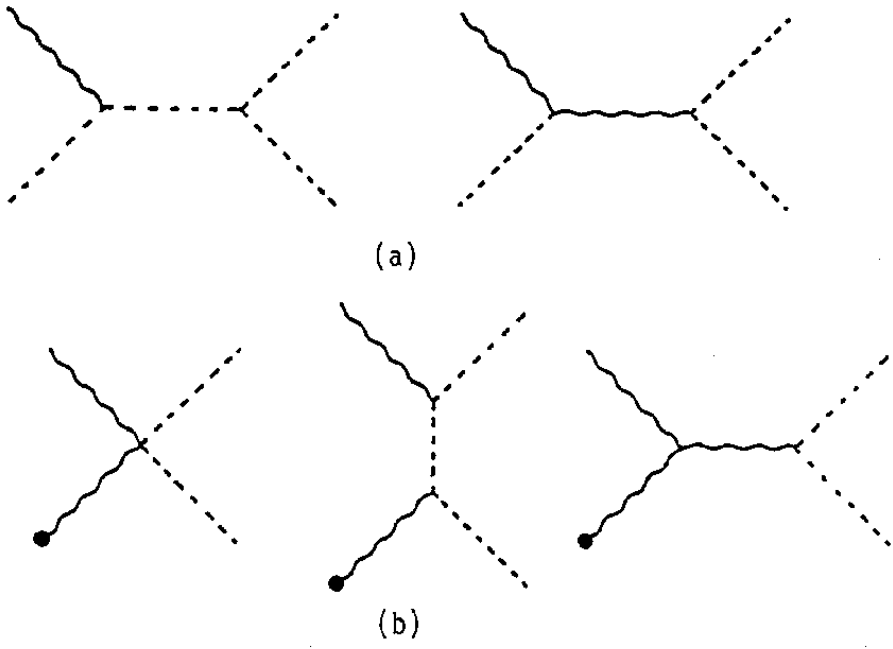
\item[Figure 4:]Some leading diagrams
for $VV\to VV$ with one transversal
and three longitudinal $V$s in the LHSM.
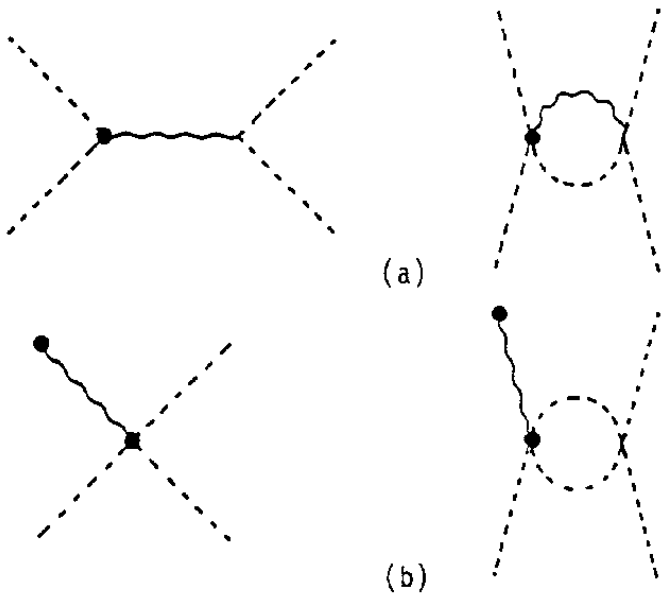
\item[Figure 5:]Some of the
tree level and one-loop diagrams which contain
the leading contribution of the terms
${\cal L}_{W\phi}$ and ${\cal L}_{B\phi}$
(\ref{lwp}) to the process $V_L V_L\to V_L V_L$. The vertex with the
dot is a nonstandard vertex, the others are standard ones.
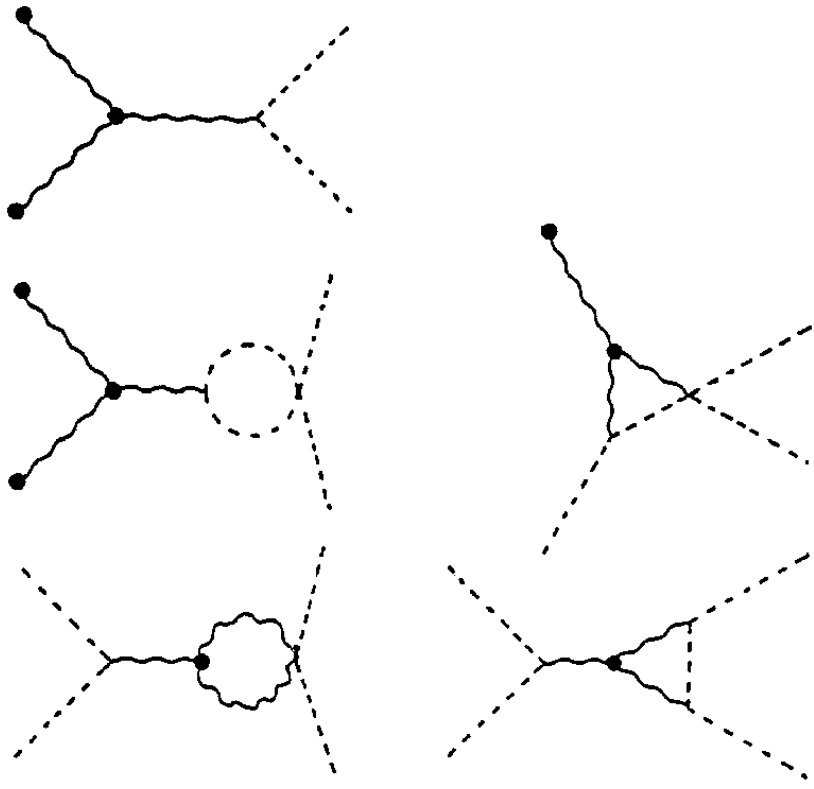
\item[Figure 6:]Same as Figure~5
for the term ${\cal L}_W$ (\ref{lw}).
\end{description}

\end{document}